# Off-site production of plasma-activated water for efficient sterilization: the crucial role of high-valence $NO_x$ and new chemical pathways

Zifeng Wang[1,2], Xiangyu Wang[2], Shenghang Xu[2], Renwu Zhou[2]*, Mingyan Zhang[2], Wanchun Li[2], Zizhu Zhang[2], Luge Wang[2], Jinkun Chen[2], Jishen Zhang[2], Li Guo[2], Dandan Pei[1], Dingxin Liu[2,3]*, and Mingzhe Rong[2]

[1] Key Laboratory of Shaanxi Province for Craniofacial Precision Medicine Research, College of Stomatology, Xi'an Jiaotong University, Xi'an 710049, China.

[2] State Key Laboratory of Electrical Insulation and Power Equipment, Center for Plasma Biomedicine, Xi'an Jiaotong University, Xi'an 710049, China.

[3] Interdisciplinary Research Center of Frontier Science and Technology, Xi'an Jiaotong University, Xi'an, 710049, China.

*Corresponding authors: renwu.zhou@xjtu.edu.cn (Renwu Zhou); liudingxin@mail.xjtu.edu.cn (Dingxin Liu)

# Abstract

Efficient sterilization of pathogens with cleaner methods is a critical concern for environmental disinfection and clinical anti-infective treatment. Plasma-activated water (PAW) is a promising alternative to chemical disinfectants and antibiotics for its strong sterilization ability and not inducing any acute toxicity, and only water and air are consumed during production. For more efficient water activation, plasma sources are commonly placed near or fully in contact with water as possible, but the risks of electrode corrosion and metal contamination of water threaten the safety and stability of PAW production. Herein, plasma-activated gas rich in high-valence $NO_x$ is generated by a hybrid plasma configuration and introduced into water for off-site PAW production. Plasma-generated $O_3$ is found to dominate the gas-phase reactions for the formation of high-valence $NO_x$. With the time-evolution of $O_3$ concentration, gaseous $NO_3$ radicals are produced behind $N_2O_5$ formation, but will be decomposed before $N_2O_5$ quenching. By decoupling the roles of gaseous $NO_3$, $N_2O_5$, and $O_3$ in the water activation, results show that short-lived aqueous species induced by gaseous $NO_3$ radicals play the most crucial role in PAW sterilization, and the acidic environment induced by $N_2O_5$ is also essential. Moreover, SEM photographs and biomacromolecule leakage assays demonstrate that PAW disrupts the cell membranes of bacteria to achieve inactivation. In real-life applications, an integrated device for off-site PAW production with a yield of 2 L/h and a bactericidal efficiency of >99.9% is developed. The PAW of 50mL produced in 3 minutes using this device is more effective in disinfection than 0.5% NaClO and 3% $H_2O_2$ with the same bacterial contact time. Overall, this work provides new avenues for efficient PAW production and deepens insights into the fundamental processes that govern the reactive chemistry in PAW sterilization.

# 1 Introduction

Pathogenic microbes coexisting with humans in most environments pose a significant threat to public health due to their proliferation and transmission [1, 2]. Spreading chemical disinfectants is one of the most effective ways to cut off the transmission of pathogens [3]. Alcohol and chlorine compounds are commonly used chemical disinfectants due to their potent sterilization effects, relatively low cost, and low pollutant loading, but injury and irritation to human eyes or skin are easily caused by their usage as a routine practice [4, 5]. Other low-irritant disinfectants, including phenolics, biguanides, and quaternary ammonium, have challenges in their self-degradation and will cause environmental enrichment and promote microbial drug resistance [6, 7]. Currently, the antimicrobials that can be directly used in human bodies are in rising demand for clinical infection control [8]. Apart from antibiotics, hydrogen peroxide is the only chemical agent widely reported for in-vivo antiseptic treatment because nontoxic byproducts will be produced [9]. However, the use of hydrogen peroxide for surgical irrigation or flushing presents a risk of mechanical injury or gas embolism [10], and peroxidase enzymes in the open wound will weaken the antiseptic effect of hydrogen peroxide [11]. Therefore, there is an immediate need to explore new alternative antimicrobial agents for environmental and medical applications.

Plasma-activated water (PAW), the aqueous solution treated by cold atmospheric plasma (CAP), has exhibited a great prospect as a green agent for eliminating pathogens [12]. Without any addition of artificially synthesized chemicals, interfacial reactions between the plasma and water can produce various reactive oxygen and nitrogen species (RONS), including $H_2O_2$, $NO_2^-$, $NO_3^-$, $O_3$, ONOOH, NO, $^1O_2$, and OH, which have been widely recognized as main contributors to the antimicrobial activity of PAW [13, 14]. These plasma-produced species are also commonly found in human physiological processes, and their activities are quite transient, allowing almost non-toxicity or environmental burden for PAW [15]. Recently, systematic safety evaluations of gavage, injection, and long-term feeding with PAW have been carried out in animal models, showing that no mortality, weight loss, organ damage, or metabolic dysfunction are observed [16-18]. Zhang *et al.* also revealed that the abdominal injection of plasma-activated saline could effectively improve survival rate, reduce bacterial load, and suppress the inflammatory response in mice suffering from sepsis [19]. Therefore, PAW can be considered safe and effective to replace conventional chemical disinfectants for environmental decontamination and clinical anti-infective treatment.

Since the first discovery by Kamgang-Youbi *et al.* about the sterilization effect of water treated with gliding arc discharge [20], numerous strategies to produce PAW for biomedical applications have been developed, with efforts in the exploration of different working gases, discharge configurations, and interaction modes between plasma and water [21]. It is generally accepted that most PAW can be efficiently produced in a short-distanced plasma-water interaction system, since some key aqueous species such as hydroxyl radicals, are mainly induced by electrons and excited particles [22]. Nevertheless, the proximity of plasma and water will reduce the robustness and consistency of the long-term operation of the water activation system. For example, the possible contact between discharge electrodes and water can lead to electrode corrosion. Therefore, in the large-scale industrial application of plasma-generated ozone for water purification, the ozone gas is introduced into the water instead of performing a discharge directly in the water [23]. In recent years, several studies have focused on the efficient preparation of PAW by combining multiple discharges or coupling plasma and microbubbles [24, 25]. Although more plasma discharges and larger plasma-water interfaces have significantly enhanced the water activation efficiency, the previous studies have not achieved the biosafety necessary for PAWs to be used for biomedical purposes, which means that the antimicrobial agents used for in-vivo flush or irrigation must be protected from contaminations. However, the in-site water activation without isolation from the plasma generator poses a risk of metal ions or discharge-generated debris getting into the water, limiting the clinical use and promotion of PAW as an antimicrobial agent. Therefore, the off-site water activation, namely introducing plasma-activated gas (PAG) into water, is more suitable for standardized PAW production, because the long-lived species in PAG allow for stable generation, online monitoring, and filtration [26]. However, short-lived gaseous species such as OH are difficult to be transferred into water by PAG, while typical long-lived gaseous species such as $N_2O$, NO, $NO_2$, and $O_3$ have less biochemical activity, limiting the efficiency of water activation.

High-valence $NO_x$ (H-$NO_x$), including gaseous $N_2O_5$ molecules and $NO_3$ radicals, are relatively stable to be used for off-site water activation, which have been recently demonstrated to have great potential for inducing the biochemical effect of PAW due to their solubility and reactivity [27, 28]. Air plasma offers convenience (at room temperature and pressure), cost (only air and electricity needed) and environment (no hazardous by-products) advantages over conventional H-$NO_x$ production methods [29, 30]. However, it is relatively challenging to produce H-$NO_x$ stably because the air plasma is very sensitive to discharge parameters and tends to switch the discharge modes [31]. Instead, the hybrid plasma discharge configuration can separately generate $O_3$ and low valence $NO_x$ (L-$NO_x$) including NO and $NO_2$, and then mix them for stable H-$NO_x$ production [32, 33]. Nevertheless, given that $NO_3$ and $N_2O_5$ usually coexist in plasma systems and are difficult to decouple, the chemical mechanisms of the gas-phase generation and liquid-phase conversion of these two species and their contribution to the sterilization of PAW have not been fully clarified. In particular, gaseous $NO_3$ radicals have not yet been quantified in plasma effluent gases due to the limitations in detection methods and their low concentration, making it difficult to understand the $NO_3$-induced reaction processes and regulate the production of $NO_3$ radicals.

In this work, to gain a deeper insight into the underlying mechanism of off-site water activation by H-$NO_x$, the total $NO_x$ concentration of PAG is set to constant, and the concentrations of $NO_3$ and $N_2O_5$ are modulated by changing $O_3$ concentration and gas temperature. A hybrid plasma discharge configuration is developed for PAG production, and a gas cell with a two-meter optical range is employed for the detection of $NO_3$ radicals. Then, the role of $NO_3$, $N_2O_5$, and $O_3$ in water activation and subsequent sterilization is distinguished by the analysis of aqueous reactive species and bactericidal effect. The key reactive species and necessary conditions for potent sterilization are also discussed. Finally, an integrated device for off-site PAW production is developed, and the sterilization effect of PAW with air or $O_2$ as the working gas of the DBD reactor is compared.

## 2 Materials and Methods

### 2.1 Experimental apparatus and setups

As shown in **Figure 1**, the hybrid plasma generation apparatus consists of a gliding arc discharge (GAD) reactor and a dielectric barrier discharge (DBD) reactor and their detailed parameters are identical to our previous study [27]. The experimental setups of the plasma generation, the measurement of electrical parameters, and the modulation and detection of gaseous reactive species are illustrated in **Figure S1**. The GAD reactor using 1.0 SLM of air as the working gas is employed to generate $NO_x$-rich gas composed of NO and $NO_2$. The discharge power of GAD is maintained at approximately 40 W. To achieve a constant total $NO_x$ concentration in the mixed PAG, the DBD reactor uses pure $O_2$ as the working gas to generate $O_3$-rich gas, so there is no DBD-generated $NO_x$ to increase the acidity of PAW. The flow rate of $O_2$ is 1 SLM and the $O_3$ concentration can be tuned by changing the discharge power of DBD from 0 W to 20 W. The voltage and current waveforms of the DBD reactor with different discharge power are shown in **Figure S2,** where the frequency of the high-voltage excitation is 10 kHz, and the peak-to-peak value of the voltage is about 10 kV when the DBD discharge power is 20 W. Since the sampling frequency of the oscilloscope is high enough, the discharge power of the DBD is derived by calculating the cycle average of the product of voltage and current. The $NO_x$-rich gas is mixed with the $O_3$-rich gas with different $O_3$ concentrations to generate mixed PAG with different compositions. To investigate the thermal decomposition of high-valence $NO_x$, the PAG can be heated by a heating wire wrapped around the outside of the quartz tube. Varying the DC voltage to drive the heating wire offers different temperature rises and the temperature of the heated PAG is measured by a digital temperature sensor (Analog Devices, Inc., DS18B20). The PAG is introduced into a gas-washing bottle containing 50 mL of deionized water for 3 minutes to produce PAW.

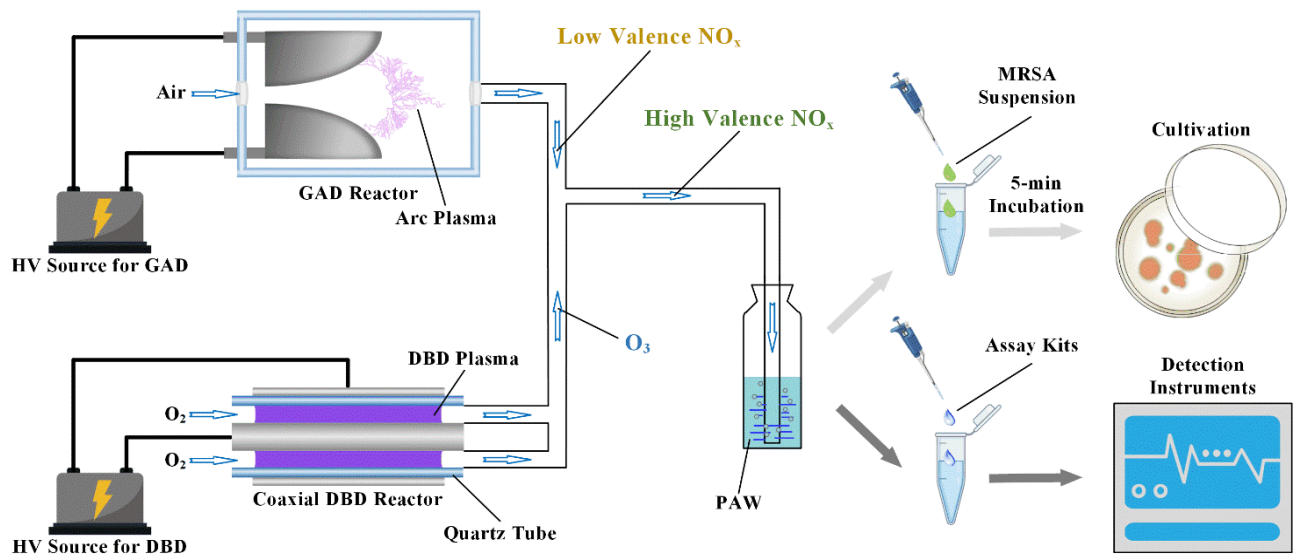

**Figure 1.** Diagram for plasma generation, water activation, reactive species measurement, and sterilization.

## 2.2 Detection of reactive species in PAG

The long-lived gaseous species including $O_3$, $N_2O$, $NO$, $NO_2$, and $N_2O_5$ in PAG are detected by a Fourier transform infrared (FTIR) spectrometer (Bruker, Tensor II) and a self-made gas cell with 0.2 m optical path. The gas cell for FTIR spectra is made of a quartz cylinder with two pieces of KBr crystal optical window. The species concentrations are quantitatively analyzed according to the Lambert-Beer law:

$$c = \frac{A}{S \times l} \qquad (1)$$

where $c$ is the absolute concentration (molecules/cm$^3$); $A$ is the area of the specific absorption peak in FTIR spectra; $S$ is the area of the corresponding standard absorption cross-section; $l$ is the optical length of the gas cell. In order to reduce measurement and calculation errors, standard NO and $NO_2$ gases are used to calibrate the FTIR system. The calibration results indicate that the FTIR system maintains good linearity for $NO_2$ in the range of 0-20,000 ppm. The standard absorption cross-sections for the gaseous species are obtained from the HITRAN database [34]. The quantitative analysis is performed using the absorption bands at the following wavenumber or wavelength: $O_3$ at ~1055 cm$^{-1}$; NO at 1900 cm$^{-1}$; $NO_2$ at 1630 cm$^{-1}$; $N_2O_5$ at ~744 cm$^{-1}$. However, FTIR absorption spectroscopy is not available for the measurement of $NO_3$ radicals, because the characteristic absorption band of NO$_3$ is in the visible range at 662 nm [35], which exceeds the detection limit of the FTIR spectrometer. In view of the concentration of plasma-generated NO$_3$ radicals as low as about 10 ppm [36], which is about 100 times lower than O$_3$ concentrations, conventional detection equipment for visible absorption spectroscopy is difficult to quantify the concentration of NO$_3$ radicals accurately. Herein, a gas cell for ultraviolet or visible (UV/Vis) absorption spectra with a 1.2 m optical path is employed. The long optical range enhances the absorption of NO$_3$ radicals and thus allows for the measurement in low concentration. A laser-driven light source (Energetiq, EQ-99) and a multi-channel spectrometer (Avantes, AVS-DESKTOP-USB2) are used for absorption spectroscopy. The standard absorption cross-section of NO$_3$ radicals is illustrated in **Figure S3.** Quantitative analysis of gaseous reactive species is calibrated by NO$_2$ standard gas (purity >99%). The concentration of gaseous reactive species is verified to remain relatively constant during water activation and optical detection.

## 2.3 Measurement of physicochemical properties and reactive species in PAW

The measurements of pH value and oxidation-reduction potential (ORP) employ a pH/ORP meter (Mettler Toledo, S210). The concentrations of H$_2$O$_2$, NO$_2^-$ and NO$_3^-$ in PAW are measured by using a microplate reader (Thermo, Varioskan Flash) with the hydrogen peroxide assay kit (Beyotime) and the nitrate/nitrite colorimetric assay kit (Beyotime). A spin trap, TEMPONE-H (1-Hydroxy-2,2,6,6-tetramethyl-4-oxo-piperidine, Enzo), is used to measure the total concentration of peroxynitrite (ONOO$^-$/ONOOH) and O$_2^-$ by an electron spin resonance (ESR) spectrometer (Bruker, EMXplus). Fifty microliters of TEMPONE-H aqueous solution with a concentration of 10 mM is mixed with 450 μL of PAW. The method of quantitative analysis for TEMPONE-H is described in a previous study [37]. Because TEMPONE-H cannot distinguish peroxynitrite and O$_2^-$, a fluorescent probe, CBA (coumarin boronic acid pinacol-ate ester, Cayman), is used to detect peroxynitrite separately. The sample for fluorescence measurement is composed of 50 μL of CBA aqueous solution with a concentration of 100 μM and 50 μL of PAW, and the sample is measured by the microplate reader after a 15-minute incubation. The concentration of aqueous O$_3$ is measured using a spectrophotometer (Hach, DR3900) with the ozone testing reagent in ampules (Hach). All experimental operations are carried out as described in the instruction manuals.

## 2.4 Assessment of sterilization effect of PAW

Methicillin-resistant *Staphylococcus aureus* (MRSA, ATCC 33591) and *Escherichia coli* (*E. coli*, ATCC 25922) are used to evaluate the sterilization effect of PAW. First, single colonies are inoculated in 4 ml of liquid medium and incubated at 37°C overnight. MRSA and *E. coli* are cultured using tryptone soy broth (TSB) and lysogeny broth (LB) culture media, respectively. Then, the cultures are transferred to fresh liquid culture media and incubated for 3 hours. Finally, the transferred cultures are centrifuged and resuspended in saline to achieve an optical density at 600 nm ($OD_{600}$) equal to ~1.0 for use as bacteria suspensions (bacterial concentration ~$10^9$ CFU/ml). The $OD_{600}$ is measured by a spectrophotometer (SHIMADZU, U-1800). Immediately after the water activation, 900 μL of PAW is mixed with 100 μL of bacteria suspension and incubated for 5 minutes. The mixture is diluted in a ten-fold gradient with PBS (phosphate buffer solution, pH = 7.4), and 10 µl of each dilution is dropped onto TSB or LB agar plates. The number of surviving bacteria is determined by counting the colonies after overnight incubation at 37 °C. The untreated water is used as a blank control.

### 2.5 Morphological changes and leakage of biomacromolecules from bacteria

The bacterial samples before and after being treated with PAW are analyzed using a compact scanning electron microscope (HITACHI, Flexsem1000II). All samples are coated with gold sputtering before scanning. In addition, a microvolume UV-vis spectrophotometer (Thermo, NanoDrop One) is used to measure the content of nucleic acids (including DNA and RNA) and proteins in the liquid samples for the assessment of biomacromolecules leakage from bacteria caused by PAW treatment. The detection wavelengths for nucleic acids and proteins are 205 nm and 260 nm, respectively.

## 3 Results and Discussions

### 3.1 Modulation of high-valence $NO_x$ generation in PAG.

The modulation of reactive species in PAG and clarification of their production and decomposition processes are fundamental for investigating the role of H-$NO_x$ in water activation. The $NO_x$-rich gas generated by GAD and the $O_3$-rich gas without $NO_x$ generated by DBD are mixed for stable generation of H-$NO_x$. **Figure S4** illustrates the FTIR absorption spectra of the PAG without $NO_x$ generation ($O_3$-rich gas mixed with air) with different discharge power of the DBD reactor. In the plasma region generated by the DBD reactor, $O_2$ molecules are dissociated into O atoms, which then react with other $O_2$ molecules to form $O_3$ (**R1**). Since pure $O_2$ is used as the working gas for DBD, only gaseous $O_3$ is generated, and higher DBD power yields more $O_3$. As shown in **Figure 2a**, when the DBD power is set at 0 W, the PAG with $NO_x$ addition, namely the $NO_x$-rich gas mixed with $O_2$, mainly consists of NO and $NO_2$. NO is generated from the combination reaction between N and O atoms dissociated from $O_2$ and $N_2$ molecules by the arc plasma. The oxygen in the air further oxidizes some NO to $NO_2$, and thus forms a coexisting mixture of $O_2$, NO, and $NO_2$. As the $O_3$ generation increases with boosting DBD power, NO is first depleted to generate $NO_2$ at a DBD power of 3W, and then almost all $NO_2$ is converted to $N_2O_5$ when the DBD power increases to 6W. With the further increased DBD power, the $N_2O_5$ detected in the FTIR spectra remains roughly constant, whereas the concentration of $NO_3$ radicals measured by UV/Vis absorption spectroscopy increases significantly, together with the $O_3$ concentration. Since the $O_2$-supplied DBD does not generate any $NO_x$, all $NO_x$ is produced by the air-supplied GAD with a constant discharge power, thus providing a constant total $NO_x$ concentration to facilitate subsequent mechanism investigations. In addition, GAD is recognized as a more efficient discharge form for generating $NO_x$, so using a hybrid plasma configuration with both GAD and DBD helps to improve the energy efficiency of high-valence $NO_x$ generation [27]. The reactions between $O_3$ and $NO_x$ are via the following pathways [38]:

$$O_{2(g)} + O_{(g)} \rightarrow O_{3(g)} \quad (R1)$$

$$NO_{(g)} + O_{3(g)} \rightarrow NO_{2(g)} + O_{2(g)} \quad (R2)$$

$$NO_{2(g)} + O_{3(g)} \rightarrow NO_{3(g)} + O_{2(g)}, \quad k_2 = 1.4 \times 10^{-19} e^{-\frac{2470}{T_g}} \quad (R3)$$

$$NO_{2(g)} + NO_{3(g)} + M \rightarrow N_2O_{5(g)} + M, \quad k_3 = 3.6 \times 10^{-42} (\frac{300}{T_g})^{4.1} \quad (R4)$$

$$N_2O_{5(g)} + M \rightarrow NO_{2(g)} + NO_{3(g)} + M, \quad k_4 = 1.3 \times 10^{-9} (\frac{300}{T_g})^{3.5} e^{-\frac{11000}{T_g}} \quad (R5)$$

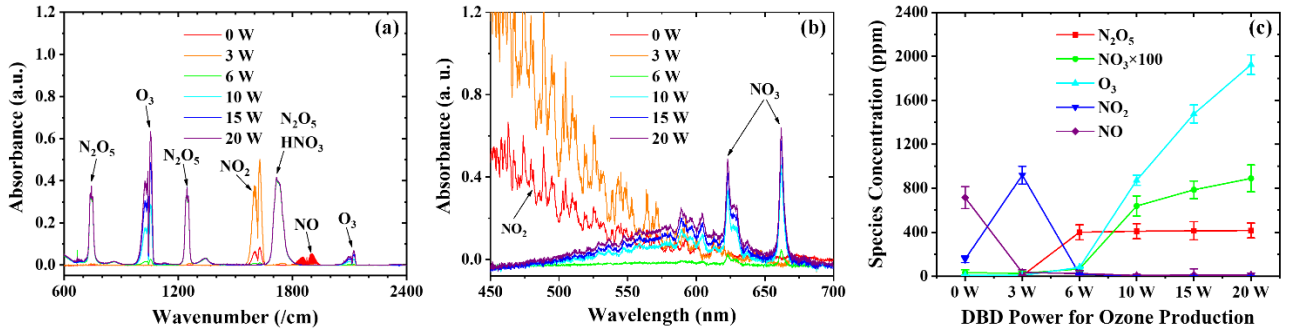

**Figure 2.** (a) FTIR absorption spectra, (b) UV/Vis absorption spectra, and (c) concentrations of $N_2O_5$, $NO_3$, $O_3$, $NO_2$, and NO in the PAG with different DBD power when the GAD power is 40 W.

**Figure 2c** shows the concentrations of $N_2O_5$, $NO_3$, $O_3$, $NO_2$, and NO in PAG with different DBD power densities. The concentration of $NO_3$ radicals increases more than 10 times from 6 W to 20 W, but it still does not exceed 10 ppm. In contrast, the concentration of $N_2O_5$ remained stable at ~400 ppm with the increase of DBD power, indicating that most of $NO_x$ are still in the form of $N_2O_5$. The above results can be explained by analyzing the rate constants of reaction **R3**-**R5**. On the one hand, when $O_3$ exists in the PAG, the equilibrium concentration of $NO_2$ is lower than that of $NO_3$ radicals due to **R3**. When $O_3$ concentration increases, the increase in the rate of **R3** ($k_2n(NO_2)n(O_3)$, where $n$ is the concentrations (molecule/m$^3$) of reactive species) induces more conversion from $NO_2$ to $NO_3$, which makes the product value of $n(NO_2)n(NO_3)$ decreases, and thus leads to a decrease in the rate of **R4** ($k_3n(NO_2)n(NO_3)n(M)$) for the $N_2O_5$ generation. These results in a trace amount of $N_2O_5$ decomposition to make up for the rate difference between **R4** and **R5**, which finally causes an increase in the concentration of $NO_3$ radicals. On the other hand, the reaction rate affects the equilibrium concentration of the reactants. Since the PAG is at room temperature (~300 K) and atmospheric pressure (1 atm), the rate of $N_2O_5$ generation is about $8.78 \times 10^{-17} n(NO_2)n(NO_3)$, and the rate of $N_2O_5$ decomposition is about $3.78 n(N_2O_5)$ according to the rate constants of **R4** and **R5**. Therefore, even if the $NO_2$ concentration in PAG is only 1 ppm (~$2.44 \times 10^{19}$ molecule/cm$^3$), the rate of $N_2O_5$ generation from $NO_3$ (**R4**) is over 500 times faster than the rate of $N_2O_5$ decomposition (**R5**) with the same initial concentrations of $N_2O_5$ and $NO_3$, resulting in the equilibrium concentration of $NO_3$ radicals being much lower than that of $N_2O_5$.

Therefore, $N_2O_5$ as the main $NO_x$ product (with small amounts of $NO_2$ and $NO_3$) can be obtained at the DBD power of 6 W, while a similar $N_2O_5$ concentration with different $NO_3$ concentrations can be regulated with different DBD power densities. By comparing the physicochemical properties and bactericidal effects of the PAW produced by these PAG conditions, it will be easy to distinguish the role of $NO_3$ and $N_2O_5$ in water activation. Moreover, since $O_3$ concentration in PAG is up to ~2000 ppm, its effect also cannot be negligible in plasma-enabled water activation.

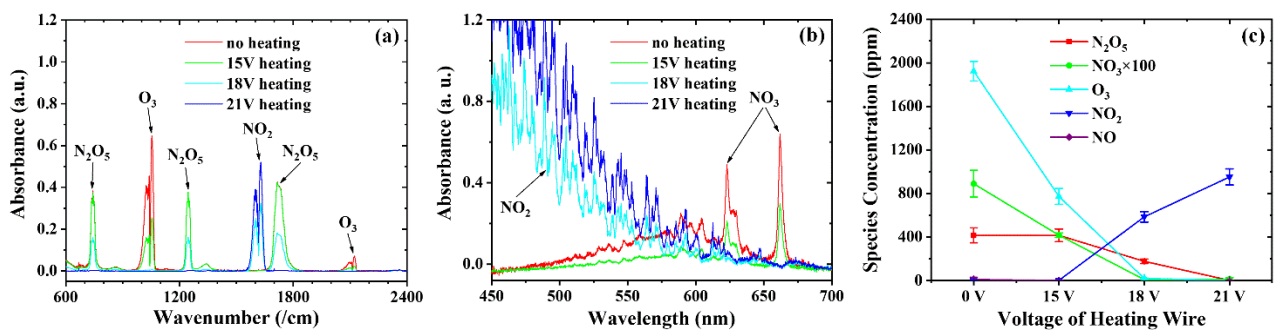

**Figure 3.** (a) FTIR absorption spectra, (b) UV/Vis absorption spectra, and (c) concentrations of $N_2O_5$, $NO_3$, $O_3$, $NO_2$, and NO in the PAG with different voltages of the heating wire. The GAD power is 40 W and the DBD power is 20 W.

To further investigate the reaction pathways of H-$NO_x$, the PAG with the DBD power of 20 W is post-heated to decompose H-$NO_x$ partially or completely through temperature regulation. The PAG temperature at the outlet of the heating tube is 68.31±0.38 °C, 78.08±0.76 °C, and 85.63±0.48 °C when the applied DC voltage of the heating wire is set at 15 V, 18 V, and 21 V, respectively. As shown in **Figure 3**, heating H-NO$_x$ is basically the inverse process of increasing O$_3$ concentration. With 15 V heating, O$_3$ and NO$_3$ are substantially reduced, but N$_2$O$_5$ concentration remains consistent with the PAG without heating. As the heating voltage increases, the O$_3$ and NO$_3$ will completely decompose, and then N$_2$O$_5$ begins to decrease and eventually be converted into NO$_2$ completely. Although the thermal decomposition of N$_2$O$_5$ at higher temperatures will produce more NO$_3$ radicals via **R5** [39], the experimental result indicates that NO$_3$ still decomposes rapidly with an increase of temperature, which is consistent with the trend in O$_3$. Nevertheless, it cannot be fully determined whether the decrease in the concentration of NO$_3$ radicals is directly due to its thermal decomposition or indirectly due to the quenching reaction of O$_3$ at a temperature over 60 °C [40]. Considering the above results showing that the concentration of NO$_3$ radicals is dominated by O$_3$, it is likely that the depletion of NO$_3$ is mainly attributed to the quenching reaction of O$_3$. In addition, the decomposition of N$_2$O$_5$ is also reported to occur when O$_3$ is depleted [27], but the decrease of N$_2$O$_5$ is later than that of NO$_3$ and O$_3$. Therefore, the chemical process of H-NO$_x$ in the gas phase is mainly dominated by O$_3$: the trend of NO$_3$ radicals is almost identical to that of O$_3$, while NO$_3$ radicals are generated slower than N$_2$O$_5$ but decomposed faster than N$_2$O$_5$ with the variation of O$_3$ concentration.

## 3.2 Chemical mechanism of water activation by high-valence NO$_x$

NO$_3$ and N$_2$O$_5$ are highly soluble and reactive with water, which is considered to play a major role in water activation. However, the specific chemical mechanisms and pathways of each species in the reactions with water have not been fully clarified. As described before, when the DBD power exceeds 6 W, N$_2$O$_5$ concentration in PAG remains almost constant, whereas the concentration of NO$_3$ radicals increases significantly with O$_3$. Therefore, the contribution of NO$_3$ and N$_2$O$_5$ in producing aqueous reactive species can be easily distinguished. Moreover, the O$_3$ concentration is also an important variable in PAG, so the PAG without NO$_x$ production is also used for water activation, and the physicochemical properties of PAW and the concentrations of short-lived aqueous species are compared to investigate the role of O$_3$. **Figure 4a** illustrates the pH and ORP values of PAW with different DBD power without NO$_x$ addition (GAD power = 0 W). In view of the absence of NO$_x$, the pH value is barely reduced, suggesting that the amount of reactive nitrogen species in PAW is negligible. In contrast, the ORP value exhibits a significant increasing trend with the increase of DBD power, which can be attributed to the dissolution of more ROS from the PAG.

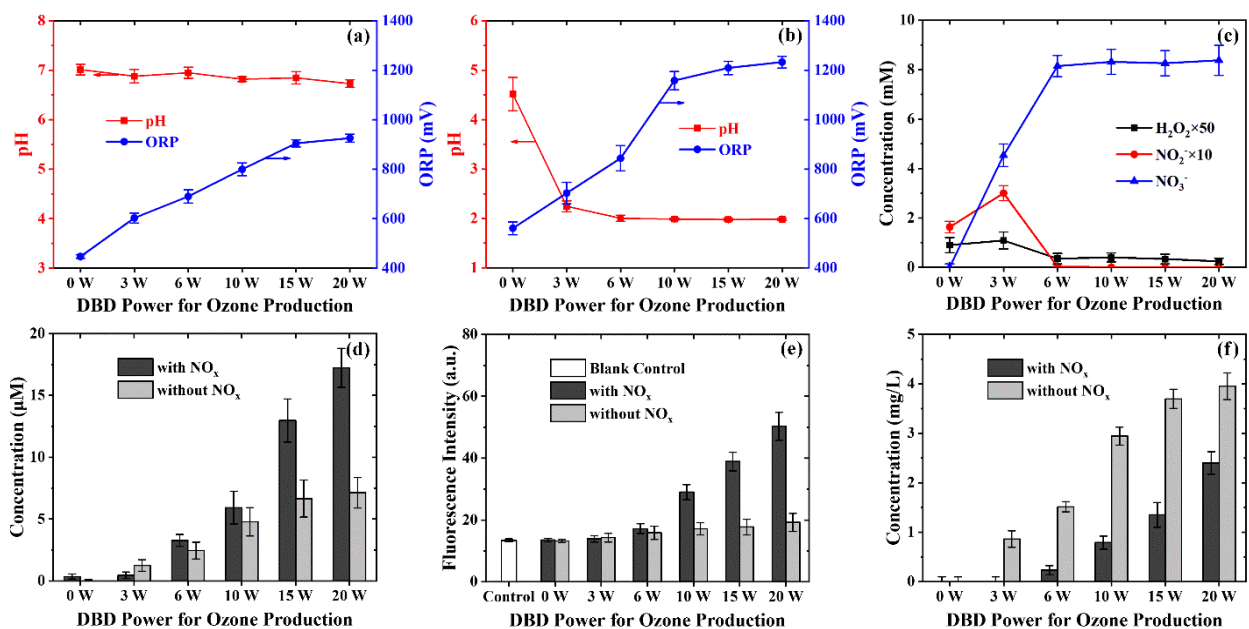

**Figure 4.** The pH and ORP values of the PAW with different DBD power when the GAD power is (a) 0 W and (b) 40 W; (c) concentrations of long-lived aqueous species including $H_2O_2$, $NO_2^-$, and $NO_3^-$ in the PAW with different DBD power when the GAD power is 40 W; (d) total concentration of peroxynitrite and $O_2^-$, (e) relative concentration of peroxynitrite reflected by the fluorescence of COH, (f) concentration of aqueous $O_3$ in the PAW with different DBD power when the GAD power is 40 W (with $NO_x$) and 0 W (without $NO_x$).

For the PAG with $NO_x$ addition, as shown in **Figure 4b**, both the acidity and ORP value of PAW rise as the DBD power increases from 0 W to 6 W, but different trends occur with the DBD power from 6 W to 20 W, where the ORP value continues to rise above 1200 mV while the pH value keeps constant about 2.0. This result can be interpreted as the acidifying capacity of PAG is limited by the total $NO_x$ concentration, when almost all the oxidized nitrogen in the PAG is converted to $N_2O_5$ and $NO_3$, which in turn form $H^+$ and $NO_3^-$ in water. The $NO_3^-$ concentration shown in **Figure 4c** also exhibits the same trend as the solution pH. However, the increased $O_3$ and $NO_3$ concentrations in PAG with boosting DBD power can continually enhance the oxidation capacity of PAW. In view of the remarkable impact of acidity on the sterilization effect of PAW [41], almost the same pH value would be helpful for subsequent investigation of sterilization. **Figure 4c** illustrates the concentrations of the long-lived species in the PAW. Compared with the results shown in **Figure 2c**, it is obvious that the concentration trends of $NO_2^-$ and $NO_3^-$ in PAW are basically the same as those of $NO_2$ and $N_2O_5$ in PAG, which is consistent with our previous studies [27]. This result implies that the gaseous $NO_2$ and $N_2O_5$ are the main precursors for producing aqueous $NO_2^-$ and $NO_3^-$. However, no gaseous reactive species with a similar concentration trend to aqueous $H_2O_2$ is found. The only source of aqueous $H_2O_2$ in PAW is the dissolution of gaseous $H_2O_2$, since no short-lived gaseous species can react with water after the tube delivery. Since there is almost no water vapor in the cylinder gas, the trace amount of gaseous $H_2O_2$ in PAG is not enough to be detected by FTIR absorption spectroscopy due to its absorption cross-section overlaps with other gaseous species such as $N_2O_5$, $HNO_3$, and $N_2O$, which also results in the $H_2O_2$ concentration in PAW not exceeding 20 μM. The long-lived $NO_2^-$ and $NO_3^-$ species in PAW are produced via the following chemical pathways [38, 42]:

$$2NO_{2(g)} + H_2O \rightarrow NO_{3\,(aq)}^- + H^+_{(aq)} + HNO_{2(aq)} \quad (R6)$$

$$HNO_{2(aq)} \leftrightarrow NO_{2\,(aq)}^- + H^+_{(aq)} \quad (R7)$$

$$N_2O_{5(g)} + H_2O \rightarrow 2NO_{3\,(aq)}^- + 2H^+_{(aq)} \quad (R8)$$

Short-lived aqueous species, including peroxynitrite, $O_2^-$, and $O_3$, have all been reported to be responsible for the sterilization effect of PAW [43-45], so their concentrations are measured herein. As shown in **Figure 4d**, the total concentration of peroxynitrite and $O_2^-$ in the PAW produced by PAG both with and without the presence of $NO_x$ exhibits an increasing trend with the increased DBD power, which is consistent with the trend of gaseous $O_3$ and $NO_3$ concentrations in PAG. The PAG without $NO_x$ inside only contains $O_3$ species, and the $O_3$ concentration can reach about 4000 ppm with the DBD power of 20 W. In contrast, the PAG with $NO_x$ production contains $O_3$ of ~2000 ppm and $NO_3$ radicals of ~9 ppm at the same total discharge power. However, the PAG with $NO_x$ addition produces much more peroxynitrite and $O_2^-$ in PAW than the PAG without $NO_x$ addition, suggesting that $NO_3$ radicals play a much more significant role than $O_3$ in inducing the production of short-lived aqueous species. The relative concentration of peroxynitrite shown in **Figure 4e** has a highly similar trend with the concentration of $NO_3$ radicals in the PAG with $NO_x$ addition, and almost no peroxynitrite is detected in the PAW without gaseous $NO_x$ addition. This result further confirms the direct correlation between gaseous $NO_3$ radicals in PAG and aqueous peroxynitrite in PAW. **Figure 4f** illustrates that the concentration of aqueous $O_3$ in PAW has an increasing trend with the DBD power, and the aqueous $O_3$ concentration in the PAW without gaseous $NO_x$ addition is much higher than that in the PAW with $NO_x$ addition, which exhibits the same trend as the gaseous $O_3$ in PAG (see **Figure 2a** and **Figure S4** for comparison). This result implies that the dominant production pathway of aqueous $O_3$ is the solution of gaseous $O_3$. Since the concentration of $N_2O_5$ in the PAG is constant with the DBD power from 6 W to 20 W, while the concentrations of peroxynitrite, $O_2^-$, and aqueous $O_3$ in PAW increase by several times with the increase of DBD power, the contribution of $N_2O_5$ to the production of short-lived aqueous species can be considered negligible. In addition, it has been reported that peroxynitrate ($O_2NOO^-/O_2NOOH$), which is also highly biologically active, is likely present in PAW [46]. However, due to the limitations of the detection technology, it cannot be confirmed whether the reaction between gaseous $NO_3$ and water can produce peroxynitrate, so the exact short-lived aqueous species induced by $NO_3$ need to be further investigated. Peroxynitrite, $O_2^-$, and aqueous $O_3$ in PAW are produced via the following chemical pathways [35, 39]:

$$NO_{3(aq)} + H_2O \rightarrow OH_{(aq)} + HNO_{3(aq)} \quad (R9)$$
$$OH_{(aq)} + NO_{2(aq)} \rightarrow ONOOH_{(aq)} \quad (R10)$$
$$ONOOH_{(aq)} \leftrightarrow ONOO^-_{(aq)} + H^+_{(aq)} \quad (R11)$$
$$ONOO^-_{(aq)} \leftrightarrow O_2^-_{(aq)} + NO_{(aq)} \quad (R12)$$
$$O_{3(g)} \rightarrow O_{3(aq)} \quad (R13)$$

In summary, the specific roles of the main components of the H-$NO_x$-rich PAG, including $NO_3$, $N_2O_5$, and $O_3$, in producing aqueous reactive species can be completely distinguished: $NO_3$ radicals accounts for the production of short-lived aqueous species such as peroxynitrite; $N_2O_5$ mainly produces $H^+$ and $NO_3^-$ to provide an acidic environment for PAW; the solution of gaseous $O_3$ contributes to the production of aqueous $O_3$.

## 3.3 Role of high-valence $NO_x$ and key aqueous species in PAW sterilization

During the direct plasma-liquid interactions, some hard-to-detect species, especially short-lived species produced at the gas-liquid interface, may also play an important role in the plasma activation process; hence the detailed plasma chemistries in the PAW production are difficult to be fully clarified. Fortunately, for the off-site PAW production, it is relatively easy to distinguish the role of gaseous species in the sterilization effect of PAW, because the species in PAG are relatively long-lived and detectable, which is essential for deepening the understanding of plasma biochemistry and facilitating the modulation of water activation. Also, the investigation of the key species in PAW sterilization is necessary. MRSA is a common Gram-positive pathogen, which may cause a variety of difficult-to-treat infectious diseases in clinical practice due to its high pathogenicity and multidrug resistance [47]. As a Gram-negative bacterium, *E. coli* is often used to evaluate the effectiveness of disinfection [48]. Therefore, MRSA and E. coli are employed as the target strains for sterilization experiments.

**Figure 5a** illustrates the sterilization effect of the PAW produced by the PAG with $NO_x$ addition (GAD power = 40 W) as a function of the DBD power. While the $N_2O_5$ concentration in PAG and the concentrations of $H_2O_2$, $NO_2^-$, $NO_3^-$, and $H^+$ in PAW are essentially constant with the DBD power varying from 6 W to 20 W, the numbers of surviving bacteria decrease from nearly $10^5$ CFUs (MRSA) and $10^3$ CFUs (*E. coli*) at 6 W to zero survival at 20 W. Similar results can also be found in **Figure 5b** which illustrates the sterilization effect of the PAW produced by the PAG shown in **Figure 3**. The heating of PAG leads to the decomposition of $O_3$ and H-$NO_x$, and thus significantly weakens the bactericidal performance of PAW. In particular, the gaseous $N_2O_5$ concentration at a heating voltage of 15 V is the same as the no-heating case, but the sterilization effect of the thus-produced PAW is obviously reduced. Despite previous studies claiming that $N_2O_5$ is potentially a key gaseous species for contributing to PAW disinfection [27, 49], this work demonstrates that $N_2O_5$ in PAG and long-lived species in PAW are not the main substantial contributors to the potent bactericidal effect. Instead, the sterilization effect of PAW exhibits a similar trend with the concentrations of gaseous $NO_3$ and $O_3$ in PAG and the concentrations of aqueous peroxynitrite, $O_2^-$ and $O_{3(aq)}$ in PAW, but the specific contribution of these reactive species to the sterilization requires further clarifications.

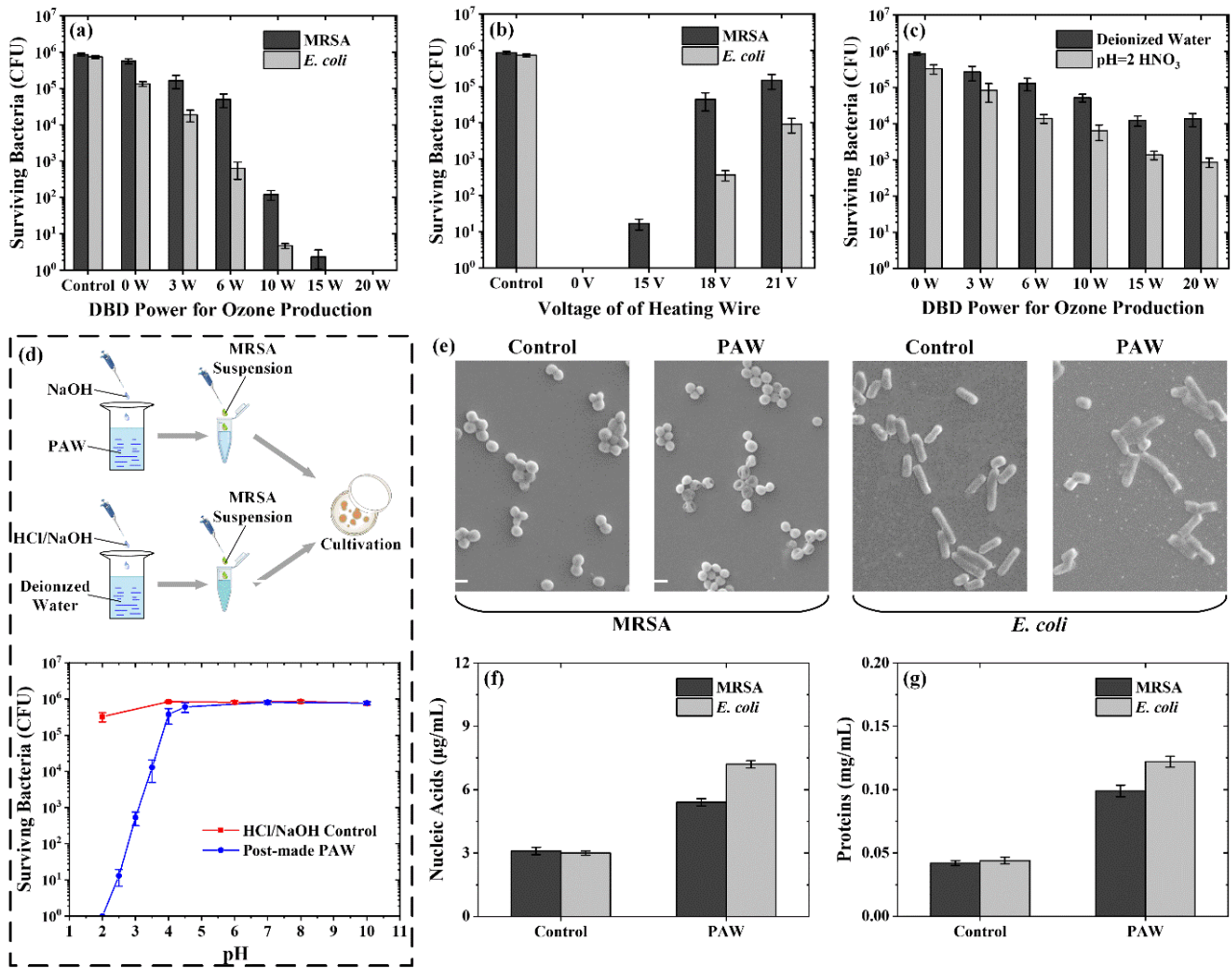

**Figure 5.** Sterilization effects of (a) the PAW with different DBD power when the GAD power is 40 W, (b) the PAW with different voltages of the heating wire, and (c) the PAW with different DBD power using deionized water and $HNO_3$ solution (pH = 2) as the solutions to be activated when the GAD power is 0 W (MRSA used only); (d) experimental method and sterilization effect (MRSA used only) of the HCl/NaOH solution and the PAW adjusted to different pH values; (e) SEM photographs of bacteria (scale bar = 1 nm) and the (f) nucleic acids (including DNA and RNA) and (g) proteins leaking from bacteria before and after being treated by PAW. The pH value of PAW is adjusted by adding NaOH solution (pH = 13) after the plasma activation. The GAD power and DBD power in figures (b), (d), (e), (f), and (g) are 40 W and 20 W, respectively.

To investigate the role of aqueous $O_3$ in the sterilization of PAW, the PAGs without $NO_x$ addition (GAD power = 0 W) with different DBD power densities are introduced into water to produce PAW. In addition, considering that $N_2O_5$ produces almost all $H^+$ and $NO_3^-$ in PAW, the $HNO_3$ solution (pH = 2) is also used for the treatment by gaseous $O_3$ to explore the synergistic water activation effect of gaseous $O_3$ and $N_2O_5$ but without $NO_3$ radicals. As shown in **Figure 5c**, the $O_3$-activated water and $HNO_3$ solution exhibit a much weaker bactericidal effect than the PAW when the GAD power is 40 W and the DBD power is 20 W. As shown in **Figure 4f**, the aqueous $O_3$ concentration in the PAW shown in **Figure 5c** will be higher than that shown in **Figure 5a**. It suggests that the strong sterilization effect of PAW cannot be achieved without the short-lived aqueous species induced by gaseous $NO_3$ radicals even with a higher aqueous $O_3$ concentration and the same concentrations of $H^+$ and $NO_3^-$. Because only $NO_3$, $N_2O_5$, and $O_3$ exist in the PAG, the crucial role of gaseous $NO_3$ radicals in sterilization can be confirmed by clarifying the minor contributions of $N_2O_5$ and $O_3$ in water activation and sterilization. Peroxynitrite and $O_2^-$, the short-lived aqueous species induced by gaseous $NO_3$ radicals, can be considered as the key aqueous species in PAW sterilization. Therefore, increasing the $O_3$ concentration in the $O_3$-$NO_x$ coexistence system can increase the concentration of $NO_3$ radicals, and thus-produced PAW has a relatively high sterilization effect.

However, the $N_2O_5$-induced acidic environment may also have a potential effect on sterilization. To verify this speculation, the NaOH solution (pH = 13) is used to counteract acidification from $N_2O_5$, which is a typical experimental method that has been employed before [50, 51]. The PAW is produced at the GAD power of 40 W and the DBD power of 20 W. As shown in **Figure 5d**, the NaOH solution is added drop by drop after the plasma activation to adjust the pH of PAW to a defined value, and the pH adjustment process is completed as quickly as possible in less than a minute. HCl/NaOH solutions in different pH values are used as the control group. The results indicate that the bactericidal effect of PAW decays rapidly with the increasing pH value, suggesting that an acidic environment is necessary for the potent sterilization effect of PAW. This can be interpreted as the increasing pH can substantially change the chemical equilibrium and alter the composition of the aqueous reactive species [52, 53]. In summary, gaseous $NO_3$ radicals play the most major role in PAW sterilization by inducing short-lived aqueous species such as peroxynitrite, while the acidic environment provided by gaseous $N_2O_5$ contributes relatively minor but essential to the bactericidal effect of PAW.

To analyze the mechanism of bacteria inactivation by PAW, the bacterial morphology before and after being treated by PAW is captured using SEM. As shown in **Figure 5e**, both PAW-treated MRSA and *E. coli* exhibit significant morphological changes, with cell membranes appearing obviously pitted and wrinkled. Particularly for *E. coli*, some white dots appear beside the bacterial cells after PAW treatment, which are likely to be biomacromolecules or organelles that leaked outside the bacteria. **Figures 5f and 5g** illustrate the content of nucleic acids and proteins in the aqueous solution leaking from the bacteria, which further demonstrates the damaging effect on bacterial cell membranes by PAW, thus leading to leakage of intracellular biomolecules. It is worth noting that *E. coli* has significantly higher nucleic acid and protein leakage than MRSA after being treated by PAW, which may be attributed to the cell wall made of peptidoglycan on the surface of the Gram-positive bacteria that somewhat resists the damage from the reactive species in PAW. Also, this is considered to be the main reason why PAW inactivates Gram-negative bacteria more effectively than Gram-positive bacteria [47]. Overall, the mechanisms of plasma-generated $NO_x$ on the PAW production and subsequent sterilization are shown in **Figure 6**.

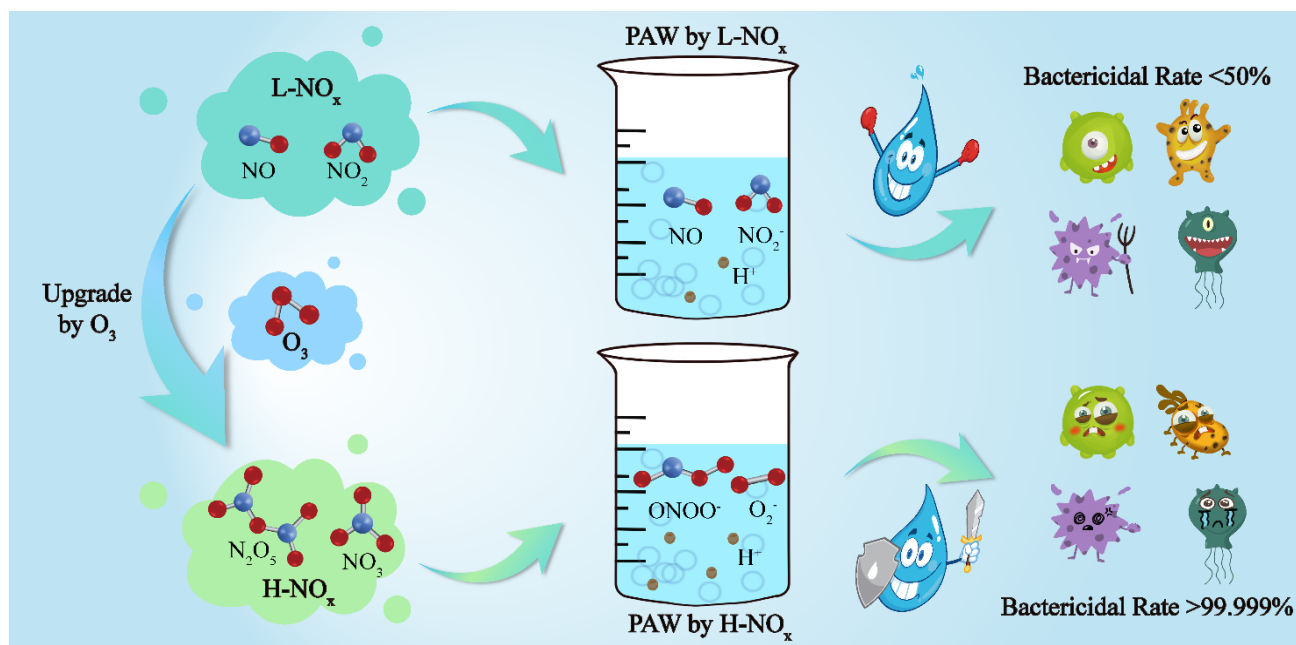

**Figure 6.** Mechanisms of the low valence $NO_x$ (L-$NO_x$) including NO and $NO_2$ and the high-valence $NO_x$ (H-$NO_x$) including $N_2O_5$ and $NO_3$ on the PAW production and subsequent sterilization.

## 3.4 Development of integrated PAW device and assessment of its sterilization effect

The equipment for standardized production of PAW is of great significance to realize the application of the novel green antimicrobial agent for environmental disinfection and clinical treatment. By comparing with previous examples of PAW production using hybrid plasma discharge configurations [27, 33], it is obvious that the use of pure $O_2$ instead of air as the working gas for the DBD in this paper can substantially improve the production efficiency of the PAW. With the same bactericidal rate of >99.999% for *S. aureus*, it takes 10 minutes to prepare 20 mL of PAW using air-supplied DBD at a total discharge power of 30 W (100 mL of PAW consumes 25 W·h of electricity) [27]; whereas this work only takes 3 minutes to prepare 50 mL of PAW using $O_2$-supplied DBD at a total discharge power of 60 W (100 mL of PAW consumes 6 W·h of electricity). Therefore, to further enhance the PAW production efficiency with the hybrid plasma discharge configuration, it is proposed to employ the molecular sieve pressure swing adsorption (PSA) module to realize portable $O_2$ production for efficient PAW production. Based on the experimental apparatus in this work, as shown in **Figure 7a**, an integrated device for off-site PAW production is developed. **Figure S5** illustrates the schematic of the integrated PAW device and the electrical characteristics of plasma reactors.

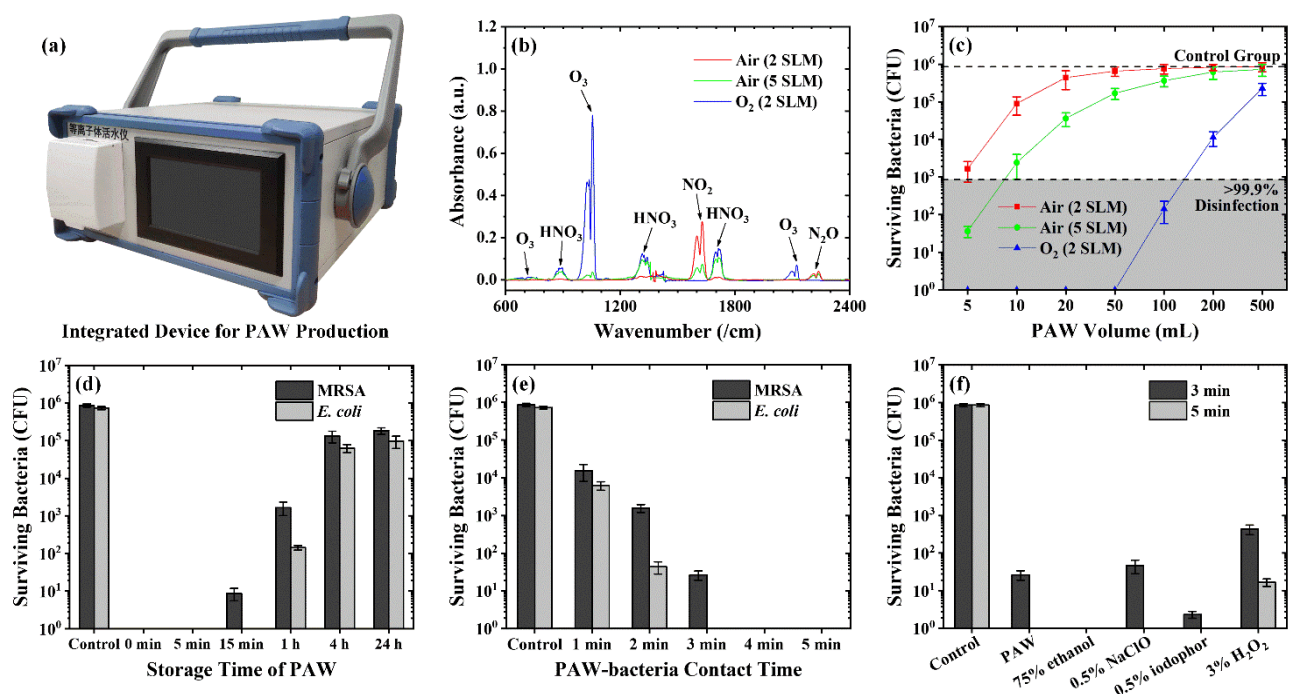

**Figure 7.** (a) Physical photo of the integrated device for off-site PAW production, (b) FTIR spectra of the PAG generated by the device using PSA-generated $O_2$ and ambient air as the working gas for the DBD reactor, (c) sterilization effect of the PAW with different volumes and different DBD working gases produced by the integrated device in 3 minutes (MRSA used only), the effects of (d) storage time and (e) PAW-bacteria contact time on the sterilization effects of the PAW prepared by the integrated device when the DBD working gas is $O_2$ at 2 SLM and the PAW volume is 50 mL, (f) comparison of sterilization effect between PAW and other disinfectants with different contact times. The default activation time and PAW-bacteria contact time are 3 and 5 minutes, respectively.

To illustrate the importance of high-purity $O_2$ in H-$NO_x$ generation and water activation, the ambient air with flow rates of 2 SLM and 5 SLM is used as the alternative working gas of the DBD reactor for comparison with high-purity $O_2$, which is highly similar to the previous hybrid plasma discharge configuration using air-supplied DBD [27, 33]. **Figure 7b** shows the FTIR spectra of the PAG generated by the integrated device with different working gases and flow rates of DBD. Since this device uses ambient air instead of the air in cylinders, the ambient water vapor converts all H-$NO_x$ into $HNO_x$ in the gas phase, including gaseous $HNO_3$ and potentially ONOOH and $O_2NOOH$. Therefore, no significant gaseous $N_2O_5$ or $NO_3$ is detected in the PAG. It is found that DBD supplied by 2 SLM of high-purity $O_2$ has a much higher $O_3$ yield, oxidizing all L-$NO_x$ to H-$NO_x$ and their derived $HNO_x$ with a large amount of excess $O_3$ in the PAG. By contrast, the air-supplied DBD with a gas flow rate of 2 SLM cannot produce enough $O_3$ to oxidize all the GAD-generated L-$NO_x$ to H-$NO_x$, so a noticeable amount of $NO_2$ is observed in the PAG. When the gas flow rate of air-supplied DBD is increased to 5 SLM, the PAG can barely achieve the coexistence of $O_3$ and $NO_2$. Considering that the $O_2$ proportion in the air is only ~21%, it is easy to understand that the lower $O_2$ concentration of the air results in less $O_3$ yield at the same power [54], because more energy will be consumed in dissociating nitrogen molecules and generating $NO_x$ that will also consume the generated $O_3$.

The sterilization effects of PAW produced by this device with different volumes are evaluated. As shown in **Figure 7c**, the bactericidal effect of PAW decreases with increasing volume since the concentrations of aqueous reactive species are getting lower simultaneously. When using PSA-generated $O_2$ for the DBD reactor, the activation time of 3 minutes can inactivate nearly 6 logs of bacteria when the PAW volume is less than or equal to 50 mL and achieves a bactericidal rate of over 99.9% with the PAW volume of 100 mL (2 L/h), which is recognized to be sufficient for most scenarios for clinical antiseptic treatment. This result also indicates that the hydration reaction of H-$NO_x$ that occurs in the gas phase does not affect the water activation effect of the PAG. When using 5 SLM of ambient air for the DBD reactor, only the PAW with 5 mL volume can reach the level of >99.9% disinfection with 3-minute activation, while the sterilization effects of the PAW with higher volumes are much weaker than the PAW using PSA-generated $O_2$. The PAW prepared with 2 SLM of air for DBD exhibits the weakest sterilization effect. This result can be attributed to the substantial difference in concentrations of $NO_3$-induced short-lived aqueous species caused by the different $O_3$ yields between air and high-purity $O_2$. Therefore, in the hybrid plasma configuration combining $NO_x$ mode and $O_3$ mode air discharges, the use of high-purity $O_2$ to feed the DBD reactor can further improve the PAW yield by about 20 times, and also increase the energy efficiency of PAW production by about 6.67 times with the energy cost of about 90 Wh/L for PAW production, taking into account the power of 120 W consumed by the PSA module and air compressor. In addition, the strong correlation between the $NO_3$-dominated chemistry in the liquid phase and the potent sterilization effect of PAW is confirmed again.

To simulate real-world scenarios, the sterilization effects of the PAW prepared by the integrated device are investigated with different storage times and bacterial contact times. As shown in **Figure**

**7d**, the PAW retains a bactericidal rate of >99% after 1 hour of storage but becomes almost completely ineffective after 4 hours of storage. As shown in **Figure 7e**, 1 minute of contact with the bacterial suspension results in ~99% sterilization, while 4 and 3 minutes of contact can achieve complete inactivation of MRSA and *E. coli*, respectively. Therefore, the contact time (similar to the D value) required for PAW to achieve 90% bacterial reduction under the default experimental condition is less than 1 minute. Compared to other disinfectants shown in **Figure 7f**, PAW exhibits better bactericidal effects than 0.5% NaClO and 3% $H_2O_2$ but not as good as 75% ethanol and 0.5% iodophor with the same contact time (3 minutes), which are commonly used disinfectant concentrations for medical and environmental applications.

**Table 1.** Comparison of the present work with previous studies on PAW production and sterilization effect.

| Plasma form | Plasma-water contact form | Plasma power | PAW yield | Sterilization effect | Ref. |
|---|---|---|---|---|---|
| Transient arc | Plasma in contact with water | 120 W | 500 mL in 20 min | 8.3 logs reduction of *S.epidermidis* with 2-5 min PAW treatment | [55] |
| Bubble + spark plasmas | Plasma in contact with water | 42-49 W | 500 mL in 30 min | >6 logs reduction of *E. coli* with 30 sec PAW treatment | [24] |
| Microbubble plasma | Plasma in contact with water | 62.1 W | 200 mL in 5 min | 2.43 logs reduction of *E. coli* with 1 min PAW treatment | [56] |
| DC-driven air plasma jet | Plasma in contact with water | Not reported | 5 mL in 5 min | >6 logs reduction of *E. coli* with 15 min PAW treatment | [46] |
| DBD plasma | PAG introduced into water (fluoride present in PAW [57]) | ~100 W | 200 mL in 7.5 min | >6 logs reduction of MRSA with 5 min PAW treatment | [26] |
| Air plasma jet + Venturi tube | PAG introduced into water | 41.4 W | | No biological effects assessed and 5-fold activation efficiency increase claimed | [58] |
| DBD+GAD plasmas | PAG introduced into water | 60 W | 50 mL in 3 min | ~6 logs reduction of MRSA and *E. coli* with 4 and 3 min PAW treatments, respectively | This work |

Given the remarkable advantages of PAW as a green antimicrobial agent, as shown in **Table 1**, there are many other techniques and pilot plants available for the production of PAW. Although some of these approaches can achieve higher energy efficiencies in PAW production or greater bactericidal efficacy, the discharge plasma in contact with water can induce intense physical fields and chemical effects that may lead to contamination of PAW with the electrodes or structural materials, thus posing a safety risk when used in the human body. Therefore, the vast majority of other approaches can only be used for environmental or agricultural applications because these PAWs do not meet the sanitary protocols necessary for the usage in in-vivo disinfection or anti-infective treatment. Based on the device using the hybrid plasma configuration presented in this work, the first randomized controlled clinical trial of PAW was completed (Registration No. ChiCTR2200056562), which demonstrates that topical pharyngeal uses of PAW significantly shortened the duration of severe acute respiratory syndrome coronavirus 2 (SARS-CoV-2) infection and alleviated patients' symptoms [59]. In addition, several clinical trials on nebulized PAW inhalation therapy (No. ChiCTR2300078706, No. ChiCTR2200066466) have been registered in the China Clinical Trial Registry (www.chictr.org.cn) and are currently under investigation. Therefore, the strategy of off-site PAW production via PAG using a hybrid plasma configuration has outstanding potential for medical applications due to its high controllability and low risks.

## 4 Conclusion

In this work, a hybrid plasma discharge configuration is developed to upgrade the plasma-generated low-valence $NO_x$ (NO and $NO_2$) into high-valence states ($N_2O_5$ and $NO_3$) for more effective production of PAW as a green alternative to chemical disinfectants. It is found that $NO_3$ radicals produce most short-lived aqueous species such as peroxynitrite, while $N_2O_5$ dissolution mainly contributes to an acidic environment. With the same $N_2O_5$ concentration in PAG, a higher concentration of $NO_3$ radicals makes PAW possess a much stronger sterilization effect, but the $HNO_3$ solution with the same acidity treated by $O_3$ has a significantly weaker bactericidal effect, suggesting that the $NO_3$-induced short-lived aqueous species plays a crucial role in the sterilization effect of PAW. Nevertheless, adding alkali after water activation will substantially weaken the bactericidal effect, indicating that the acidic environment provided by $N_2O_5$ is also necessary for sterilization. In addition, both MRSA and *E. coli* treated by PAW exhibited cell membrane rupture and leakage of intracellular nucleic acids and proteins. Finally, an integrated device for off-site PAW production is developed, and it can yield PAW of 100 mL with a bactericidal rate of >99.9% within 3 minutes. The use of high-purity $O_2$ instead of air as the working gas for DBD can increase the PAW yield by approximately 2000%, further confirming that high $NO_3$ and $O_3$ concentrations in PAG are crucial to the powerful sterilization ability of PAW. The PAW produced by this device (3 minutes activation, 50 mL volume) has a better bactericidal effect than 0.5% NaClO and 3% $H_2O_2$, but not as good as 75% ethanol and 0.5% iodophor. The integrated PAW device enables safe in-vivo use while efficiently producing PAWs with potent sterilization effects, and several clinical trials using the device have already been approved. This work elaborates chemical mechanism of the off-line PAW production by H-NO$_x$ and develops an integrated device for pilot trials, facilitating the environmental and clinical applications of PAW as a novel green antimicrobial agent.

## Supplementary Material

Supplementary material includes experimental setups of plasma generation, modulation, and measurement (**Figure S1**), voltage and current waveforms of DBD reactor with different powers (**Figure S2**), standard absorption cross-section of gaseous NO$_3$ (**Figure S3**), FTIR absorption spectra of the PAG without NO$_x$ (**Figure S4**), the schematic of the integrated PAW device and the voltage and current waveforms of DBD and GAD reactors in the integrated PAW device (**Figure S5**).

## Acknowledgments

This work is supported by the National Natural Science Foundation of China (Grant No. 52150221 and 12175175) and the Fundamental Research Funds for the Central Universities.

## Data Availability

Data will be made available on request.

## References

[1] J. M. Hassell, M. Begon, M. J. Ward, et al. (2017) Urbanization and Disease Emergence: Dynamics at the Wildlife–Livestock–Human Interface, Trends in Ecology & Evolution, https://doi.org/10.1016/j.tree.2016.09.012

[2] P. K. Pandey, P. H. Kass, M. L. Soupir, et al. (2014) Contamination of water resources by pathogenic bacteria, AMB Express, https://doi.org/10.1186/s13568-014-0051-x

[3] P. M. Schneider. (2013) New technologies and trends in sterilization and disinfection, American Journal of Infection Control, https://doi.org/10.1016/j.ajic.2012.12.007

[4] C. F. Goh, L. C. Ming, L. C. Wong. (2021) Dermatologic reactions to disinfectant use during the COVID-19 pandemic, Clinics in Dermatology, https://doi.org/10.1016/j.clindermatol.2020.09.005

[5] W. A. Rutala, D. J. Weber. (1997) Uses of inorganic hypochlorite (bleach) in health-care facilities,

Clinical Microbiology Review, https://doi.org/10.1128/cmr.10.4.597

[6] M. Aranke, R. Moheimani, M. Phuphanich, et al. (2021) Disinfectants In Interventional Practices, Current Pain and Headache Reports, https://doi.org/10.1007/s11916-021-00938-3

[7] D. Basiry, N. E. Heravi, C. Uluseker, et al. (2022) The effect of disinfectants and antiseptics on co- and cross-selection of resistance to antibiotics in aquatic environments and wastewater treatment plants, Frontiers in Microbiology, https://doi.org/10.3389/fmicb.2022.1050558

[8] D. E. Fry. (2016) Topical Antimicrobials and the Open Surgical Wound, Surgical Infections, https://doi.org/10.1089/sur.2016.107

[9] D. Bailey, E. B. Rizk. (2021) Origin and Use of Hydrogen Peroxide in Neurosurgery, Neurosurgery, https://doi.org/10.1093/neuros/nyab107

[10] G. A. Zimmerman, K. I. Lipow. (2004) Pneumocephalus with neurological dericit from hydrogen peroxide irrigation - Case illustration, Journal of Neurosurgery, https://doi.org/10.3171/jns.2004.100.6.1122

[11] H. Steiling, B. Munz, S. Werner, et al. (1999) Different types of ROS-scavenging enzymes are expressed during cutaneous wound repair, Experimental Cell Research, https://doi.org/10.1006/excr.1998.4366

[12] R. Ma, G. Wang, Y. Tian, et al. (2015) Non-thermal plasma-activated water inactivation of food-borne pathogen on fresh produce, Journal of Hazardous Materials, https://doi.org/10.1016/j.jhazmat.2015.07.061

[13] H. Xu, Y. Zhu, M. Du, et al. (2021) Subcellular mechanism of microbial inactivation during water disinfection by cold atmospheric-pressure plasma, Water Research 188116513. https://doi.org/10.1016/j.watres.2020.116513

[14] R. Zhou, R. Zhou, K. Prasad, et al. (2018) Cold atmospheric plasma activated water as a prospective disinfectant: the crucial role of peroxynitrite, Green Chemistry, https://doi.org/10.1039/c8gc02800a

[15] Q. Wang, D. Salvi. (2021) Evaluation of plasma-activated water (PAW) as a novel disinfectant: Effectiveness on Escherichia coli and Listeria innocua, physicochemical properties, and storage stability, Lwt-Food Science and Technology, https://doi.org/10.1016/j.lwt.2021.111847

[16] D. Xu, Q. J. Cui, Y. Xu, et al. (2018) Systemic study on the safety of immuno-deficient nude mice treated by atmospheric plasma-activated water, Plasma Science & Technology, https://doi.org/10.1088/2058-6272/aa9842

[17] Y. Xu, S. Peng, B. Li, et al. (2021) Systematic Safety Evaluation of Cold Plasma-Activated Liquid in Rabbits, Frontiers in Physics, https://doi.org/10.3389/fphy.2021.659227

[18] V. Nastasa, A. S. Pasca, R. N. Malancus, et al. (2021) Toxicity Assessment of Long-Term Exposure to Non-Thermal Plasma Activated Water in Mice, International Journal of Molecular Sciences, https://doi.org/10.3390/ijms222111534

[19] J. Zhang, K. Qu, X. Zhang, et al. (2019) Discharge Plasma-Activated Saline Protects against Abdominal Sepsis by Promoting Bacterial Clearance, Shock, https://doi.org/10.1097/shk.0000000000001232

[20] G. Kamgang-Youbi, J. M. Herry, M. N. Bellon-Fontaine, et al. (2007) Evidence of Temporal Postdischarge Decontamination of Bacteria by Gliding Electric Discharges: Application to Hafnia alvei, Applied and Environmental Microbiology, https://doi.org/10.1128/AEM.00120-07

[21] R. Zhou, R. Zhou, P. Wang, et al. (2020) Plasma-activated water: generation, origin of reactive species and biological applications, Journal of Physics D: Applied Physics, https://doi.org/10.1088/1361-6463/ab81cf

[22] N. K. Kaushik, B. Ghimire, Y. Li, et al. (2019) Biological and medical applications of plasma-activated media, water and solutions, Biological Chemistry, https://doi.org/10.1515/hsz-2018-0226

[23] A. Hafeez, Z. Shamair, N. Shezad, et al. (2021) Solar powered decentralized water systems: A cleaner solution of the industrial wastewater treatment and clean drinking water supply challenges, Journal of Cleaner Production, https://doi.org/10.1016/j.jclepro.2020.125717

[24] K. Hadinoto, N. Rao, J. B. Astorga, et al. (2023) Hybrid plasma discharges for energy-efficient production of plasma-activated water, Chemical Engineering Journal, https://doi.org/10.1016/j.cej.2022.138643

[25] Y. Gao, M. Li, C. Sun, et al. (2022) Microbubble-enhanced water activation by cold plasma, Chemical Engineering Journal, https://doi.org/10.1016/j.cej.2022.137318

[26] Z. Wang, S. Xu, D. Liu, et al. (2021) An integrated device for preparation of plasma-activated media with bactericidal properties: An in vitro and in vivo study, Contributions to Plasma Physics, https://doi.org/10.1002/ctpp.202100125

[27] Z. Wang, M. Zhu, D. Liu, et al. (2023) $N_2O_5$ in air discharge plasma: energy-efficient production, maintenance factors and sterilization effects, Journal of Physics D: Applied Physics, https://doi.org/10.1088/1361-6463/acb65f

[28] W. Wang, Z. Liu, J. Chen, et al. (2021) Surface air discharge used for biomedicine: the positive correlation among gaseous $NO_3$, aqueous $O_2^-$/$ONOO^-$ and biological effects, Journal of Physics D: Applied Physics, https://doi.org/10.1088/1361-6463/ac2276

[29] R. P. Wayne, I. Barnes, P. Biggs, et al. (1991) The nitrate radical: Physics, chemistry, and the atmosphere, Atmospheric Environment. Part A. General Topics, https://doi.org/10.1016/0960-1686(91)90192-A

[30] B. Ouyang, M. W. McLeod, R. L. Jones, et al. (2013) $NO_3$ radical production from the reaction between the Criegee intermediate $CH_2OO$ and $NO_2$, Physical Chemistry Chemical Physics, https://doi.org/10.1039/c3cp53024h

[31] K. Liu, Y. Hu, J. Lei. (2017) The chemical product mode transition of the air DBD driven by AC power: A plausible evaluation parameter and the chemical behaviors, Physics of Plasmas, https://doi.org/10.1063/1.5004423

[32] S. Sasaki, K. Takashima, T. Kaneko. (2021) Portable Plasma Device for Electric N2O5 Production from Air, Industrial & Engineering Chemical Research, https://dx.doi.org/10.1021/acs.iecr.0c04915

[33] Z. Wang, L. Liu, D. Liu, et al. (2022) Combination of $NO_x$ mode and $O_3$ mode air discharges for water activation to produce a potent disinfectant, Plasma Sources Science & Technology,

https://doi.org/10.1088/1361-6595/ac60c0

[34] L. S. Rothman, I. E. Gordon, A. Barbe, et al. (2009) The HITRAN 2008 molecular spectroscopic database, Journal of Quantitative Spectroscopy & Radiative Transfer, https://doi.org/10.1016/j.jqsrt.2009.02.013

[35] S. Park, W. Choe, C. Jo. (2018) Interplay among ozone and nitrogen oxides in air plasmas: Rapid change in plasma chemistry, Chemical Engineering Journal, https://doi.org/10.1016/j.cej.2018.07.039

[36] E. Simoncelli, J. Schulpen, F. Barletta, et al. (2019) UV-VIS optical spectroscopy investigation on the kinetics of long-lived RONS produced by a surface DBD plasma source, Plasma Sources Science & Technology, https://doi.org/10.1088/1361-6595/ab3c36

[37] C. Chen, F. Li, H. Chen, et al. (2017) Aqueous reactive species induced by a PCB surface micro-discharge air plasma device: a quantitative study, Journal of Physics D: Applied Physics, https://doi.org/10.1088/1361-6463/aa8be9

[38] Y. Teng, Y. Lin, W. Wang, et al. (2021) The role of $O_3$ on the selective formation of nitrate and nitrite in plasma-treated water, Journal of Physics D: Applied Physics, https://doi.org/10.1088/1361-6463/ac026d

[39] R. P. Wayne. (1991) THE NITRATE RADICAL - PHYSICS, CHEMISTRY, AND THE ATMOSPHERE-PREFACE. Atmospheric Environment Part a-General Topics, https://doi.org/10.1016/0960-1686(91)90191-9

[40] T. Shimizu, Y. Sakiyama, D. B. Graves, et al. (2012) The dynamics of ozone generation and mode transition in air surface micro-discharge plasma at atmospheric pressure, New Journal of Physics, https://doi.org/10.1088/1367-2630/14/10/103028

[41] K. Oehmigen, M. Hahnel, R. Brandenburg, et al. (2010) The Role of Acidification for Antimicrobial Activity of Atmospheric Pressure Plasma in Liquids, Plasma Processes and Polymers, https://doi.org/10.1002/ppap.200900077

[42] S. Khuntia, M. K. Sinha, B. Saini. (2020) Conversion of $NO_2$ through ozonation and peroxone process in gas and aqueous phase: Finding the suitable process through experimental route, Chemical Engineering Journal, https://doi.org/10.1016/j.cej.2020.124082

[43] P. Lukes, E. Dolezalova, I. Sisrova, et al. (2014) Aqueous-phase chemistry and bactericidal effects from an air discharge plasma in contact with water: evidence for the formation of peroxynitrite through a pseudo-second-order post-discharge reaction of $H_2O_2$ and $HNO_2$. Plasma Sources Science & Technology, https://doi.org/10.1088/0963-0252/23/1/015019

[44] S. Ikawa, K. Kitano, S. Hamaguchi. (2010) Effects of pH on Bacterial Inactivation in Aqueous Solutions due to Low-Temperature Atmospheric Pressure Plasma Application, Plasma Processes and Polymers, https://doi.org/10.1002/ppap.200900090

[45] M. J. Pavlovich, H. Chang, Y. Sakiyama, et al. (2013) Ozone correlates with antibacterial effects from indirect air dielectric barrier discharge treatment of water, Journal of Physics D: Applied Physics, https://doi.org/10.1088/0022-3727/46/14/145202

[46] M. Ma, Y. Zhang, Y. Lv, et al. (2020) The key reactive species in the bactericidal process of plasma

activated water, Journal of Physics D: Applied Physics, https://doi.org/10.1088/1361-6463/ab703a

[47] L. Guo, R. Xu, D. Liu, et al. (2019) Eradication of methicillin-resistant Staphylococcus aureus biofilms by surface discharge plasmas with various working gases, Journal of Physics D: Applied Physics, https://doi.org/10.1088/1361-6463/ab32c9

[48] Y. Zhao, S. Ojha, C. M. Burgess, et al. (2021) Inactivation efficacy of plasma-activated water: influence of plasma treatment time, exposure time and bacterial species, International Journal of Food Science and Technology, https://doi.org/10.1111/ijfs.14708

[49] A. Moldgy, G. Nayak, H. A. Aboubakr, et al. (2020) Inactivation of virus and bacteria using cold atmospheric pressure air plasmas and the role of reactive nitrogen species, Journal of Physics D: Applied Physics, https://doi.org/10.1088/1361-6463/aba066

[50] J. Chen, Z. Wang, J. Sun, et al. (2023) Plasma-Activated Hydrogels for Microbial Disinfection, Advanced Science, https://doi.org/10.1002/advs.202207407

[51] B. Pang, Z. Liu, S. Wang, et al. (2022) Alkaline plasma-activated water (PAW) as an innovative therapeutic avenue for cancer treatment, Applied Physics Letters, https://doi.org/10.1063/5.0107906

[52] S. Goldstein, J. Lind, G. Merényi. (2005) Chemistry of peroxynitrites as compared to peroxynitrates, Chemical Reviews, https://doi.org/10.1021/cr0307087

[53] J. M. Régimbal, M. Mozurkewich. (2000) Kinetics of peroxynitric acid reactions with halides at low pH, Journal of Physical Chemistry A, https://doi.org/10.1021/jp9930301

[54] K. Liu, Z. Zheng, S. Liu, et al. (2019) Study on the Physical and Chemical Characteristics of DBD: The Effect of $N_2/O_2$ Mixture Ratio on the Product Regulation, Plasma Chemistry and Plasma Processing, https://doi.org/10.1007/s11090-019-09998-1

[55] A. Pemen, P. van Ooij, F. Beckers, et al. (2017) Power Modulator for High-Yiel45d Production of Plasma-Activated Water, IEEE Transactions on Plasma Science, https://doi.org/10.1109/TPS.2017.2739484

[56] C. Man, C. Zhang, H. Fang, et al. (2022) Nanosecond-pulsed microbubble plasma reactor for plasma-activated water generation and bacterial inactivation, Plasma Processes and Polymers, https://doi.org/10.1002/ppap.202200004

[57] W. Xi, L. Guo, D. Liu, et al. (2022) Upcycle hazard against other hazard: Toxic fluorides from plasma fluoropolymer etching turn novel microbial disinfectants, Journal of Hazardous Materials, https://doi.org/10.1016/j.jhazmat.2021.127658

[58] Y. Gao, M. Li, C. Sun, et al. (2022) Microbubble-enhanced water activation by cold plasma, Chemical Engineering Journal, https://doi.org/10.1016/j.cej.2022.137318

[59] X. Guo, X. Lv, Y. Wu, et al. (2024) Safety and efficacy of plasma-activated water on prolonged viral shedding of COVID-19 patients: A randomized controlled trial, Plasma Processes and Polymers, https://doi.org/10.1002/ppap.202300195